\documentclass[iop]{emulateapj}
\usepackage[utf8]{inputenc}
\usepackage[T1]{fontenc}

\shorttitle{Helical oscillations of the velocity vector in chromospheric magnetic elements}
\shortauthors{M. Stangalini et al.}

\begin{document}

\title{Polarised Kink Waves in Magnetic Elements: evidence for chromospheric helical waves}
\author{M. Stangalini\altaffilmark{1}, F. Giannattasio\altaffilmark{1,2}, R. Erd\'elyi\altaffilmark{3}, S. Jafarzadeh\altaffilmark{4}, G. Consolini\altaffilmark{2}, S. Criscuoli\altaffilmark{5}, I. Ermolli\altaffilmark{2}, S. L. Guglielmino\altaffilmark{6}, F. Zuccarello\altaffilmark{6}}
\email{marco.stangalini@inaf.it}
\altaffiltext{1}{INAF-OAR National Institute for Astrophysics, Via Frascati 33, 00078 Monte Porzio Catone (RM), Italy}
\altaffiltext{2}{INAF-IAPS National Institute for Astrophysics, Via del Fosso del Cavaliere, 100, 00133 Rome, Italy}
\altaffiltext{3}{Solar Physics \& Space Plasma Research Centre (SP2RC), School of Mathematics and Statistics, University of Sheffield, Sheffield S3 7RH, UK}
\altaffiltext{4}{Institute of Theoretical Astrophysics, University of Oslo, P.O. Box 1029 Blindern, N-0315 Oslo, Norway}
\altaffiltext{5}{NSO, National Solar Observatory, Boulder USA}
\altaffiltext{6}{Department of Physics and Astronomy, University of Catania, Via S. Sofia 78, 95125 Catania, Italy}

\begin{abstract}
In recent years, new high spatial resolution observations of the Sun's atmosphere have revealed the presence of a plethora of small-scale magnetic elements down to the resolution limit of current cohort of solar telescopes ($\sim 100-120$ km on the solar photosphere). These small magnetic field concentrations, due to the granular buffeting, can support and guide several magneto-hydrodynamics (MHD) wave modes that would eventually contribute to the energy budget of the upper layers of the atmosphere.\\
In this work, exploiting the high spatial and temporal resolution chromospheric data acquired with the Swedish 1-meter Solar Telescope (SST), and applying the empirical mode decomposition (EMD) technique to the tracking of the solar magnetic features, we analyse the perturbations of the horizontal velocity vector of a set of chromospheric magnetic elements. We find observational evidence that suggests a phase relation between the two components of the velocity vector itself, resulting in its helical motion.
\end{abstract}


\section{Introduction}
Small-scale magnetic elements (SSMEs) with diameters of the order of a few hundred km are ubiquitous in the lower solar atmosphere \citep{2010ApJ...723L.164L, 2012A&A...539A...6B, 2012NatCo...3E1315M, 2014A&A...561L...6S}. Interestingly, they play a significant role in the energy budget of the chromosphere, by acting as magnetic conduits for magneto-hydrodynamics waves \citep{DePontieu2004, 2006ApJ...648L.151J}. Indeed, under the forcing action of the photospheric convection, SSMEs are continuously pushed, pulled, advected, and dispersed over the solar surface \citep[see for instance][and references therein]{1998ApJ...495..973B, 2012ApJ...752...48C, 2011ApJ...740L..40K, 2013ApJ...770L..36G, fabio04}. At the same time, different MHD wave modes propagating along these waveguides are also excited \citep[e.g. magneto-acoustic, kink and sausage, Alfvén;][]{1978SoPh...56....5R, 1981A&A....98..155S, Edwin1983, Roberts1983, Musielak1989, 1998ApJ...495..468S, Hasan2003, Musielak2003a,  Fedun2011, 2012ApJ...755...18V, 2012A&A...538A..79N}. In this regard, many authors have reported the presence of a plethora of waves in SSMEs at a range of heights spanning over the lower solar atmosphere. In addition, it was also found that such localised magnetic structures can support the propagation of both compressive \citep[see for example][]{2012ApJ...746..183J, 2004ApJ...617..623B} and incompressible \citep{2014ApJ...784...29M} waves, as for example kink, and Alfvén waves \citep{2007Sci...318.1572E}. Indeed, \cite{2002ApJ...567L.165M} have shown the presence of long-period waves in chromospheric bright points that are not consistent with the observational signatures expected for acoustic waves but rather MHD waves.
%
  \begin{figure}
  \centering
  \includegraphics[trim=0cm 0cm 0cm 0cm, clip, width=8.cm]{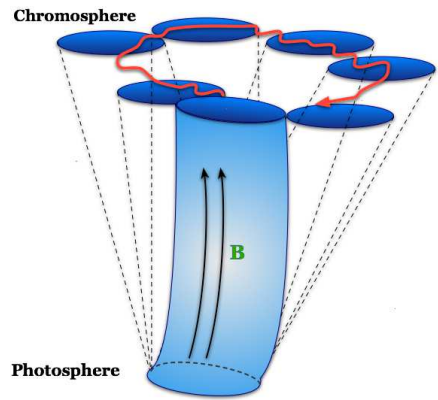}
   \caption{Cartoon of the typical displacement of a SSME as measured in the solar chromosphere. A low frequency helical displacement is superimposed on a high frequency kink-like oscillation (red line).}
    \label{Fig:cartoon}
   \end{figure} 
  \begin{figure}
  \centering
  \includegraphics[trim=0cm 0cm 0cm 0cm, clip, width=8.cm]{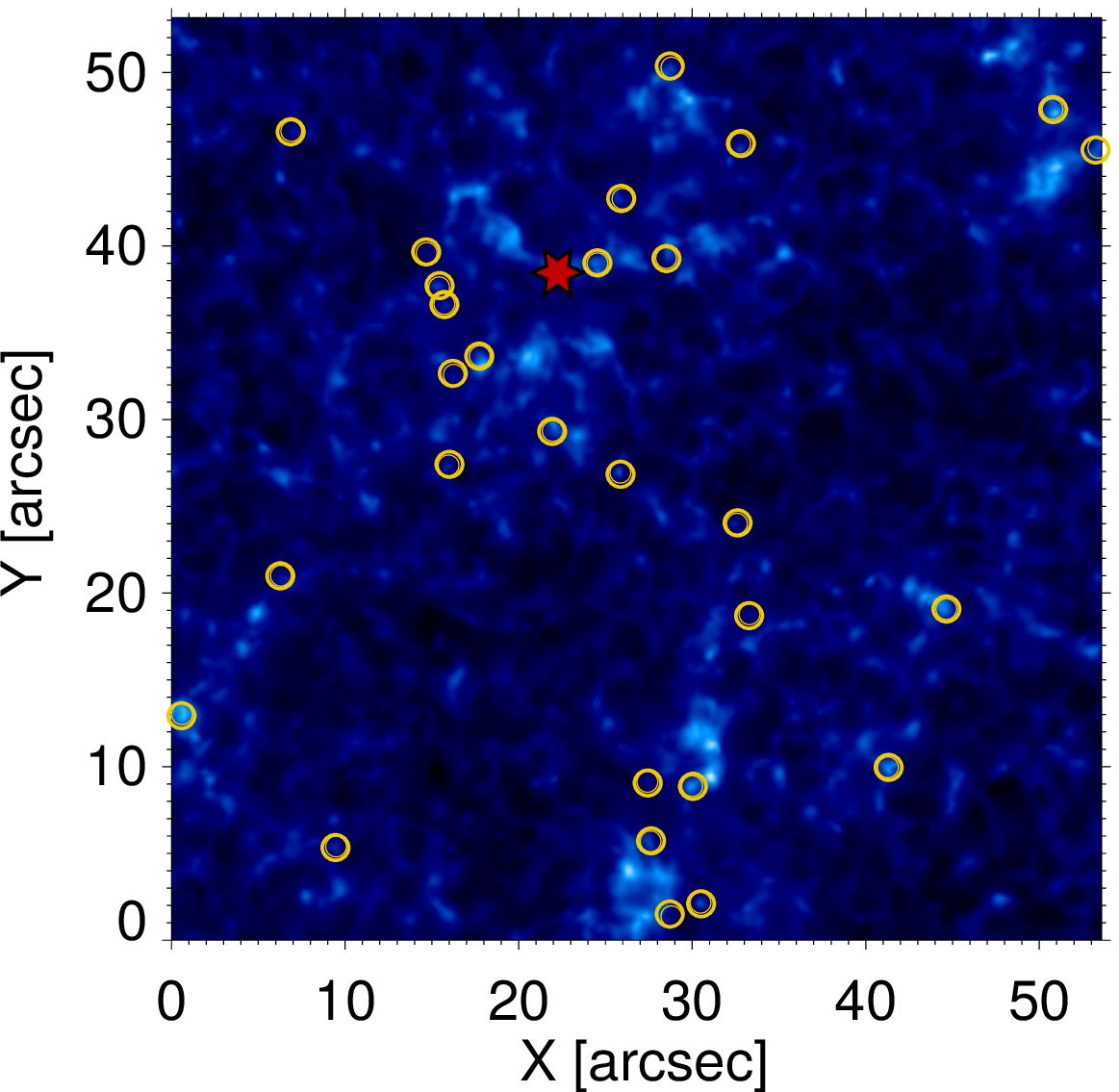}
  \includegraphics[trim=0cm 0cm 0cm 0cm, clip, width=8.cm]{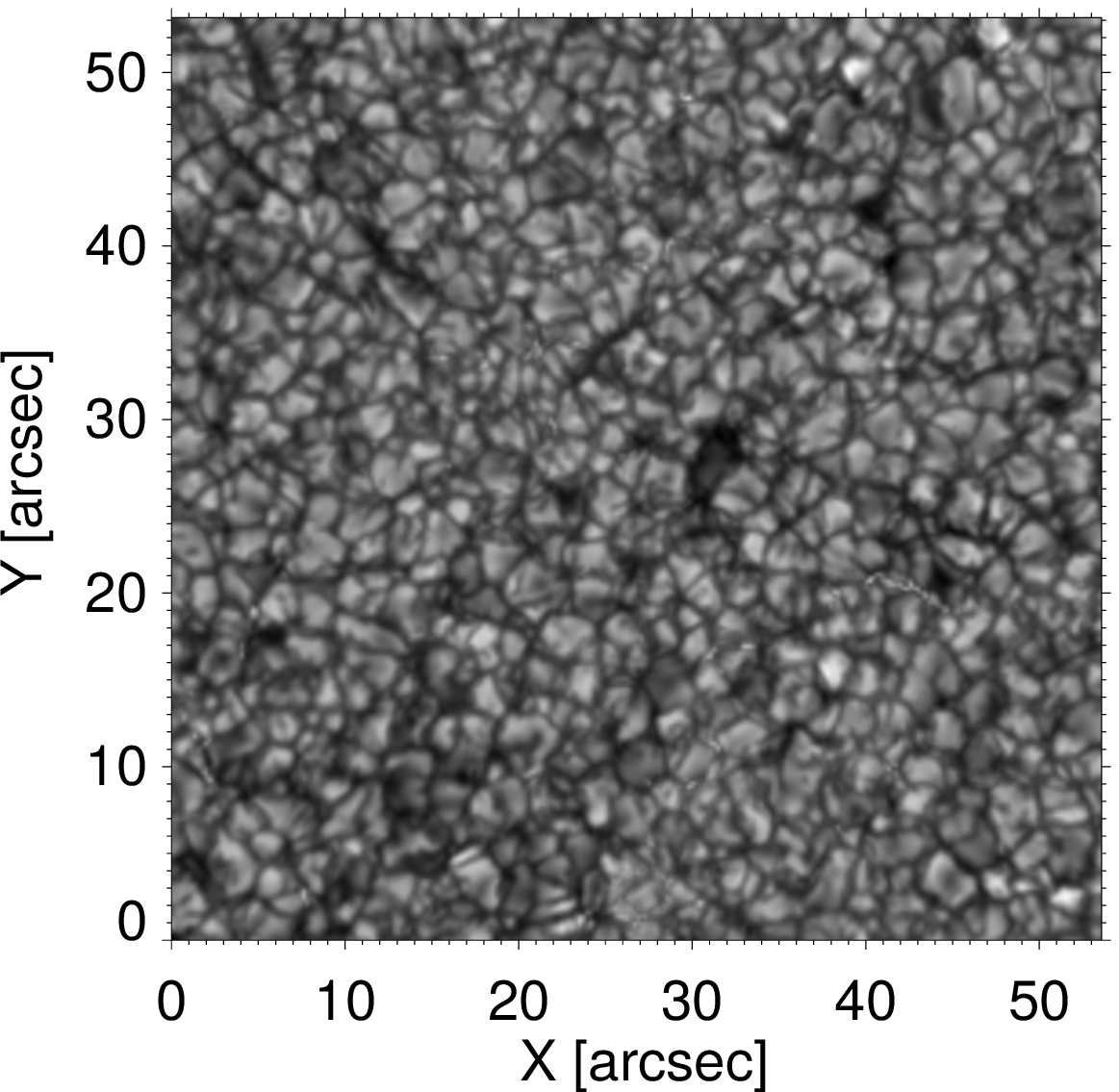}
   \caption{Example of the data analysed in this study. Top:  chromospheric Ca II H broadband image. The 35 longest-lived SSMEs used in the analysis are highlighted by yellow circles. The red star (upper panel) identifies the case study analysed in the text.} Bottom: image of the same FoV taken at 395.3 nm with $1$ \AA~ bandpass.
    \label{Fig:maps}
   \end{figure} 
Among the many types of MHD modes, kink waves have been observed at different regions in the solar atmosphere; from the lower photosphere  \citep[e.g.][]{2011ApJ...740L..40K}, to the chromosphere \citep[e.g.][]{2013A&A...549A.116J}, and the corona \citep[e.g.][]{2007Sci...317.1192T, 2011Natur.475..477M}. It is generally believed that kink waves are continuously generated thanks at least to the photospheric granular buffeting action. In this regard, an observational proof supporting this scenario was recently provided by \citet{refId0}, who observed the presence of several sub-harmonics in the kink-like oscillations of SSMEs in the photosphere, with a fundamental period that is consistent with the photospheric granular timescale. Furthermore, the reported presence of sub-harmonics can be regarded as the signature of a chaotic excitation \citep{2009arXiv0910.3570S, 2010arXiv1002.3363S}. More recently, \citet[][hereafter Paper I]{2015A&A...577A..17S}, using high-resolution simultaneous observations at different heights of the solar atmosphere, observed the propagation of kink waves from the photosphere to the chromosphere.\\ 
In this work, we move on in this field by studying the temporal orientation of the velocity vector of kink perturbations. We consider the same 35 magnetic elements analysed in Paper I, and investigate their horizontal motion. We assume that the studied features are chromospheric SSMEs. This assumption is based on the presence of circular polarization signals at their base in the photosphere, and the large coherence between the oscillatory signals in the photosphere and chromosphere reported in Paper I.

\section{Dataset and methods}
The dataset employed in this work was acquired on 2011 August 6 with the Swedish Solar Telescope \citep[SST, ][]{2003SPIE.4853..370S}. The obtained data consists of a series of chromospheric broadband images centered at the core of the Ca II H line at $396.9$ nm on a quiet Sun region at disk center. The estimated formation height of the spectral line is $700$ km above the photosphere \citep{2016arXiv161003104J}. In Paper I, the SSMEs were tracked to study kink wave propagation from the photosphere to chromosphere. The set of magnetic elements constituted a collection of the longest-lived ones for which a chromospheric counterpart was confirmed by visual inspection of Ca II H data. Therefore this work here may be regarded as a continuation of Paper I, but with a focus on the chromosphere only, where the magnetic elements are not forced directly to move around by the solar photospheric granulation and, perhaps even more importantly, they are free to oscillate.   
  \begin{figure}
  \centering
  \includegraphics[trim=0cm 0cm 0cm 0cm, clip, width=8cm]{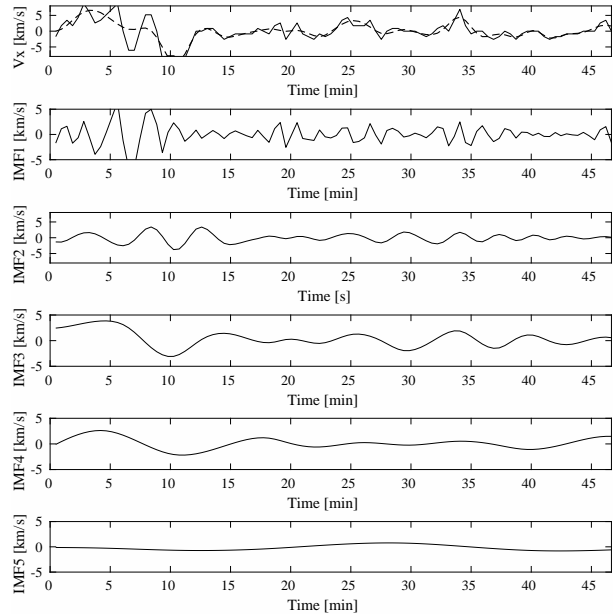}
   \caption{Example of EMD of one of the components of the horizontal velocity vector of a SMME tracked in this study. The black solid line in the upper panel represents the original time series, while the dashed line the filtered signal. The subsequent panels show the IMFs that decompose the signal. }
    \label{Fig:EMD}
   \end{figure}

   \begin{figure*}
  \centering
  \includegraphics[trim=0cm 0cm -1cm 0cm, clip, width=7cm]{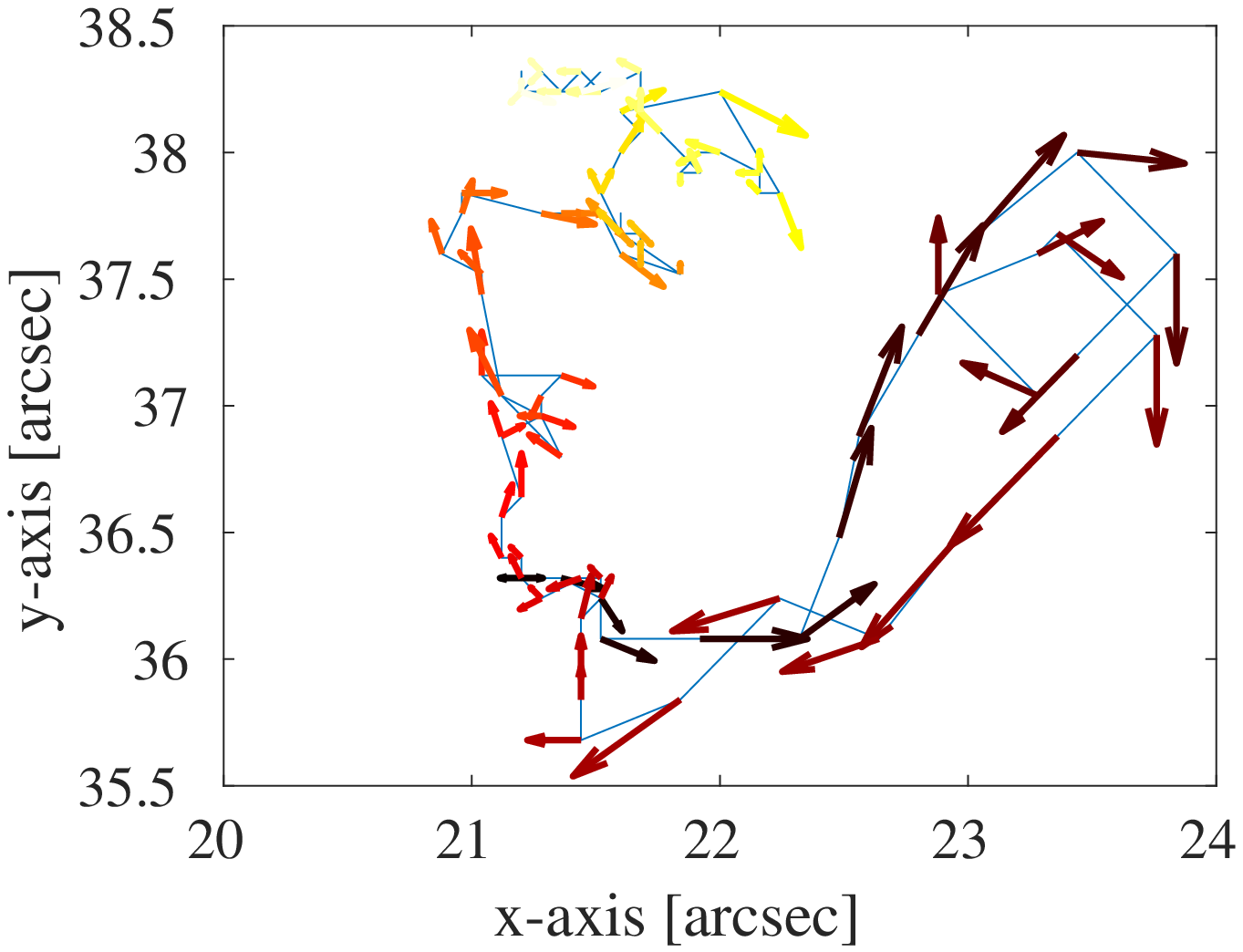}
  \includegraphics[trim=0cm 0cm -1cm 0cm, clip, width=7.5cm]{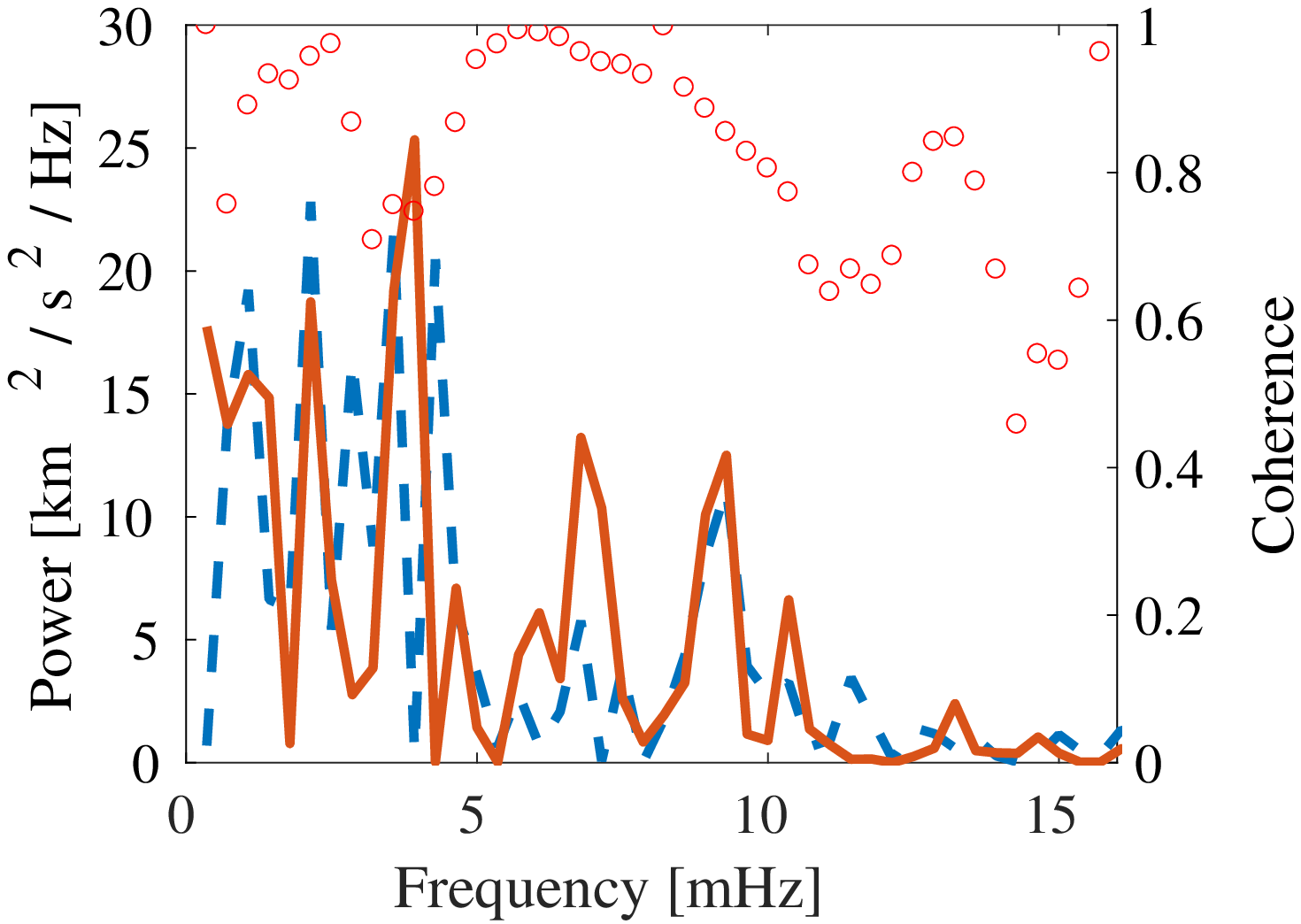}\\
   \caption{\textit{Left panel }: Trajectory in time $\vec{s}(t)$ of the magnetic element marked with a red star in Fig. \ref{Fig:maps}, with the velocity vector $\dot{\vec{s}}(t)$ superimposed. The colors encode the temporal evolution, from dark red to light yellow. \textit{Right panel:} Periodogram of the two components of $\dot{\vec{s}}(t)$ ($v_{x}$ red line, $v_{y}$ dashed blue line). In the same plot, we also show with red circles the coherence spectrum between these two components, smoothed with an averaging window $3$ points wide.}
    \label{Fig:elem24}
   \end{figure*}

\begin{figure*}
  \centering
  \includegraphics[trim=0cm -1cm 0cm -2cm, clip, width=5cm]{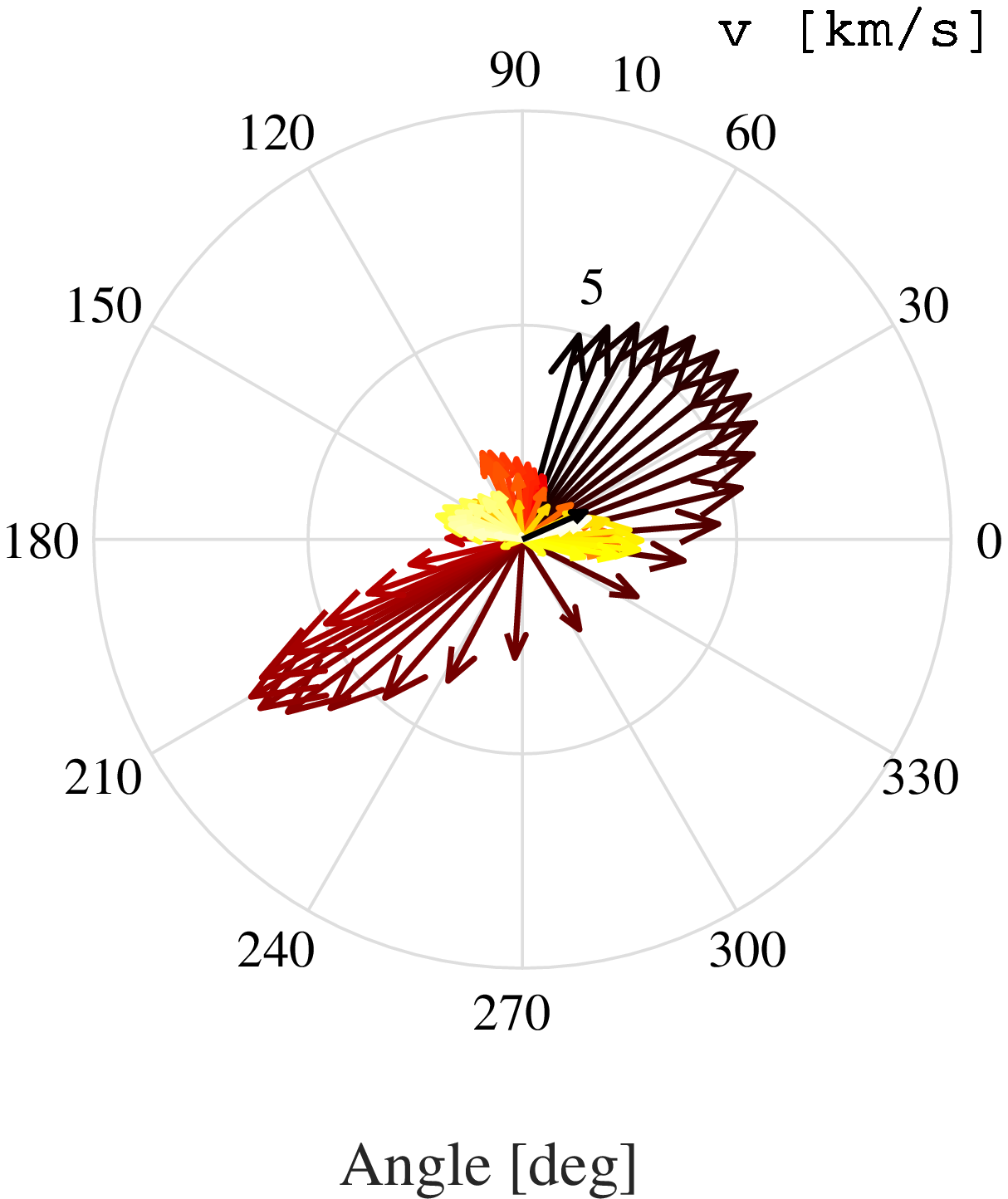}
  \includegraphics[trim=-1cm -1cm 0cm 0cm, clip, width=7cm]{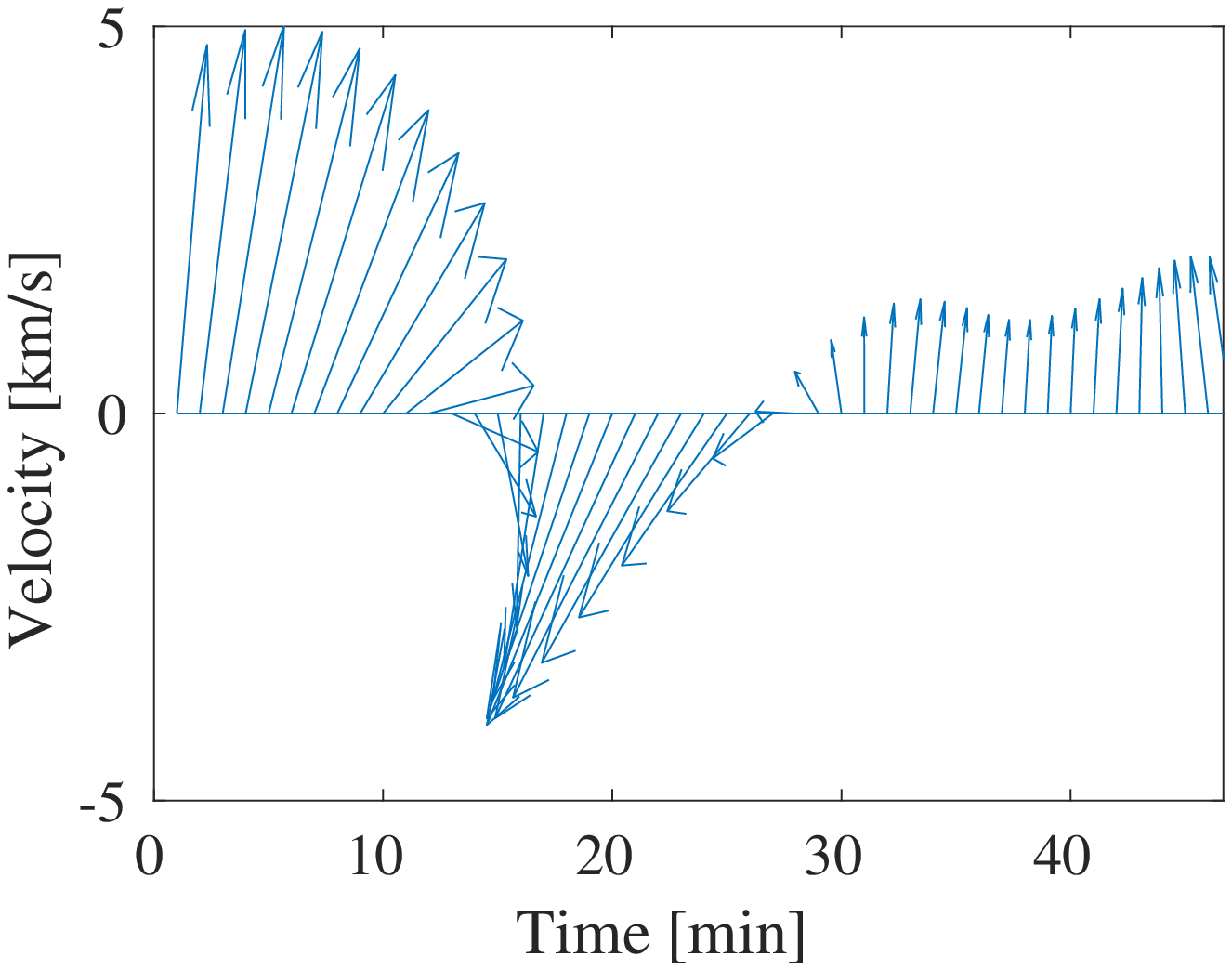}
   \caption{\textit{Left panel }: Evolution of the velocity vector $\dot{\vec{s}}(t)$ of the element marked as a star in Fig. \ref{Fig:maps} as a function of time, in polar coordinates. The color scale encodes the temporal evolution from dark red to yellow. The latitudinal circles represent the magnitude of the vector in $km/s$. \textit{Right panel:} Velocity vector (orientation and magnitude) as a function of time.}
    \label{Fig:elem24compass}
   \end{figure*}

   \begin{figure*}
  \centering
  \includegraphics[trim=0cm 0cm 0cm 0cm, clip, width=5.cm]{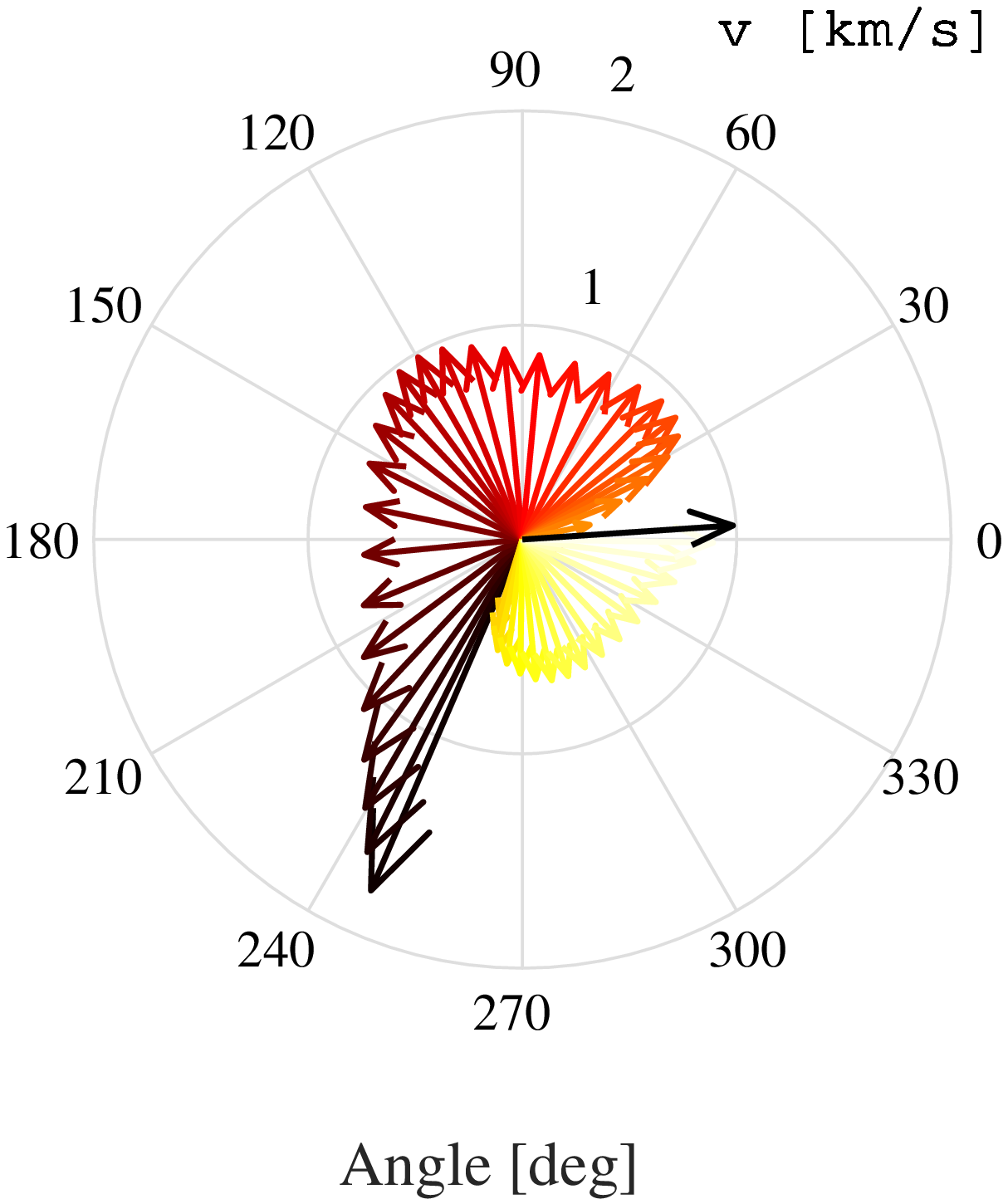}
  \includegraphics[trim=0cm 0cm 0cm 0cm, clip, width=5.cm]{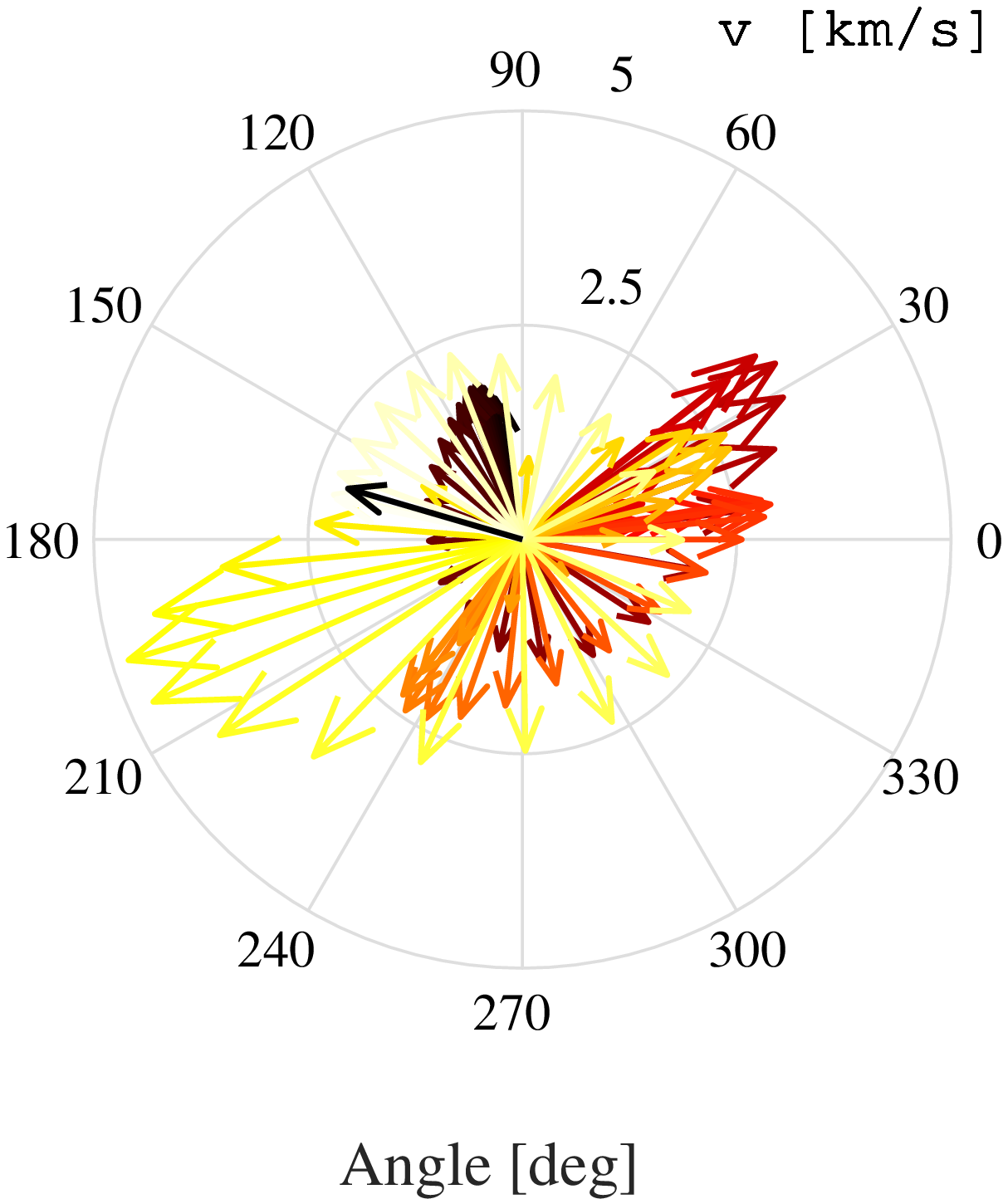}
  \includegraphics[trim=0cm 0cm 0cm 0cm, clip, width=5.cm]{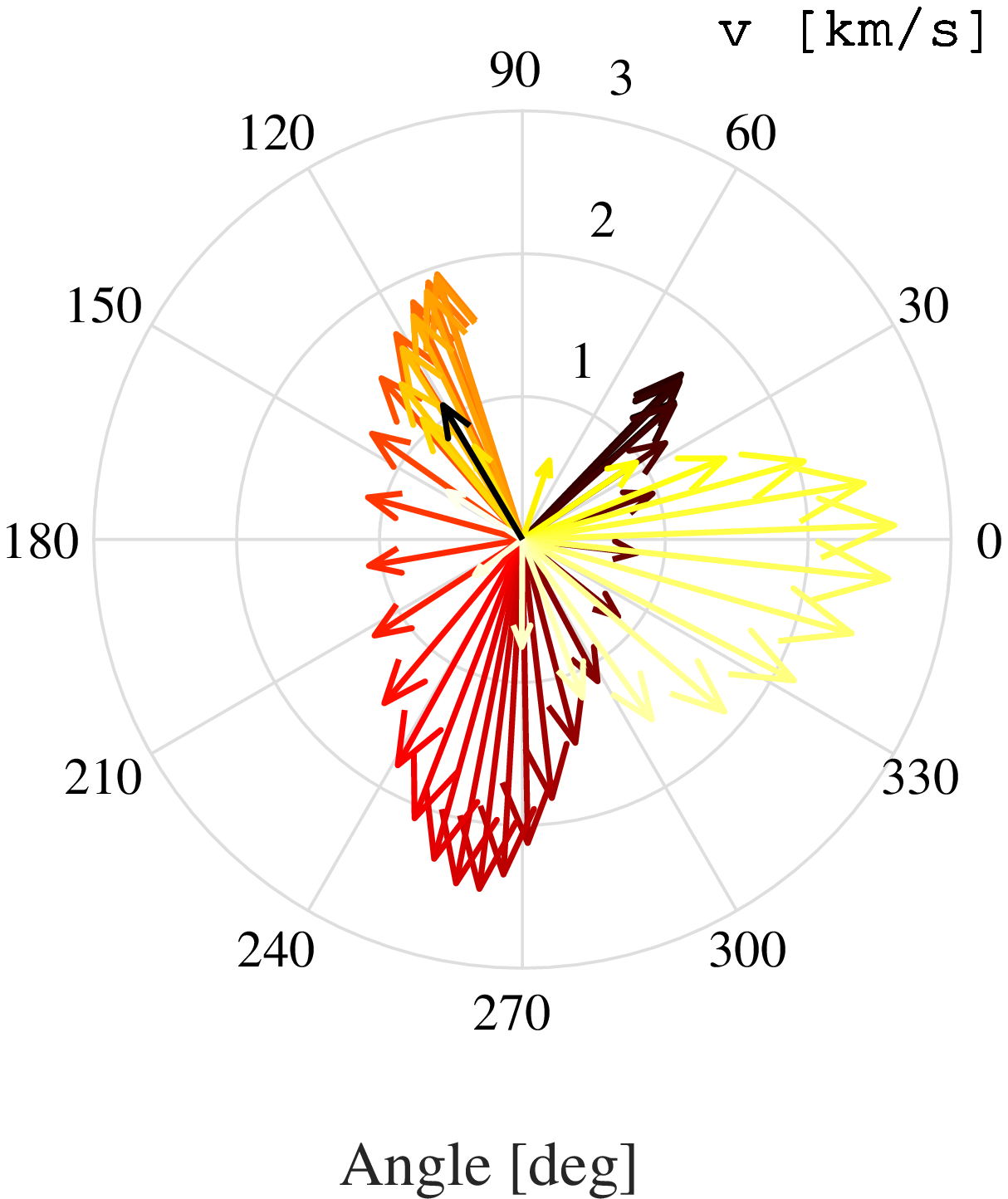}\\
  \vspace{0.7cm}
  \includegraphics[trim=0cm 0cm 0cm 0cm, clip, width=5.cm]{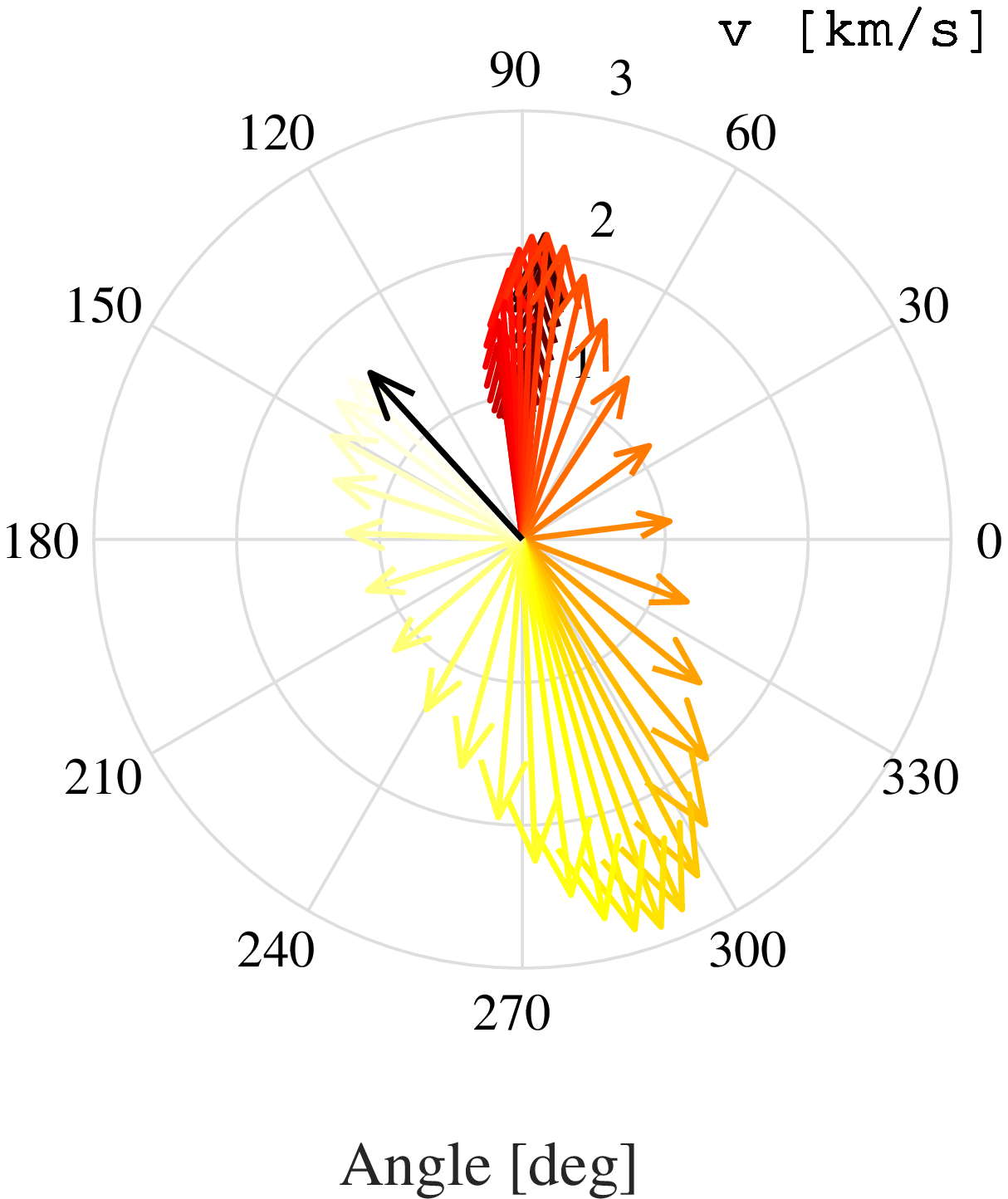}
  \includegraphics[trim=0cm 0cm 0cm 0cm, clip, width=5.cm]{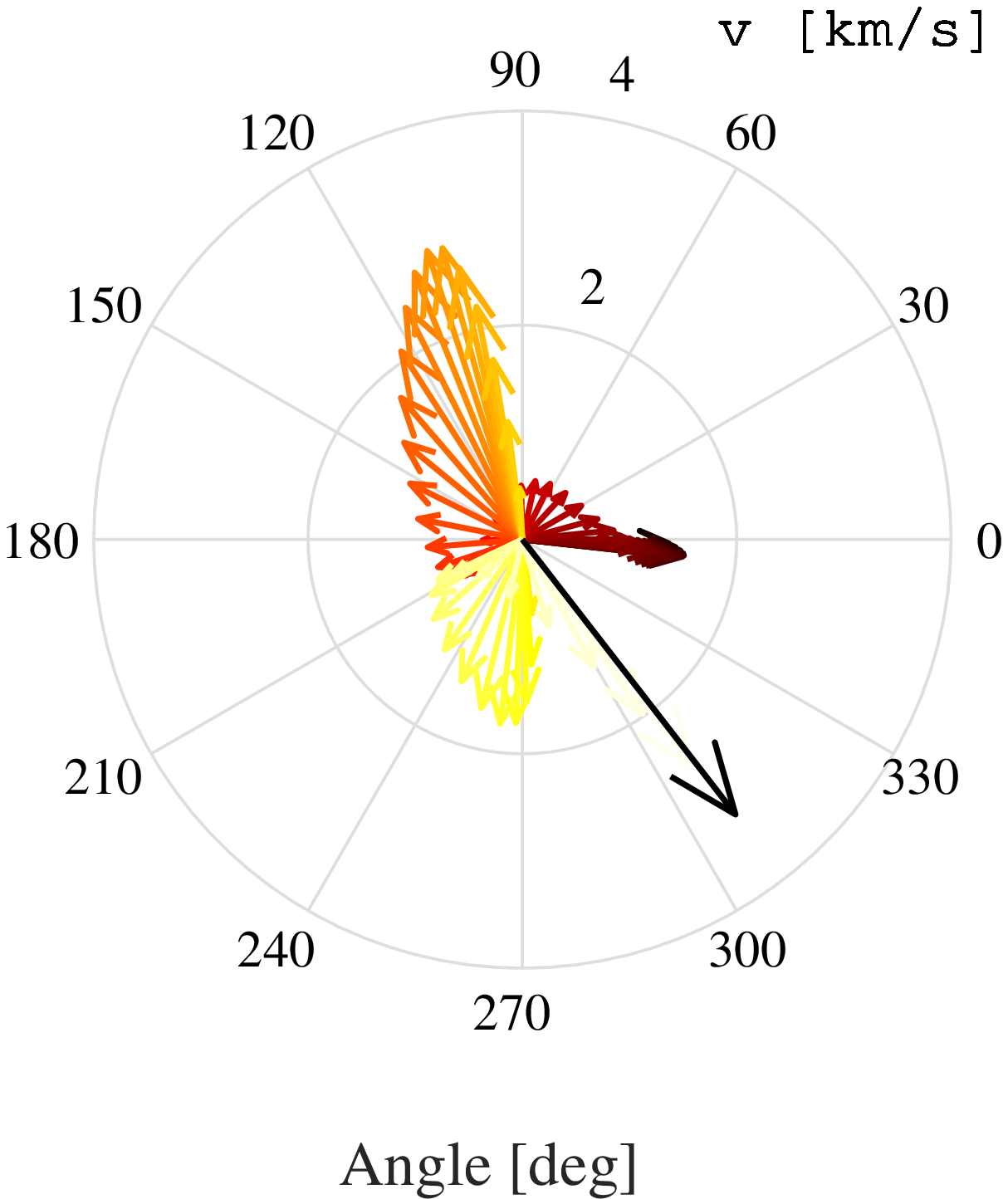}
  \includegraphics[trim=0cm 0cm 0cm 0cm, clip, width=5.cm]{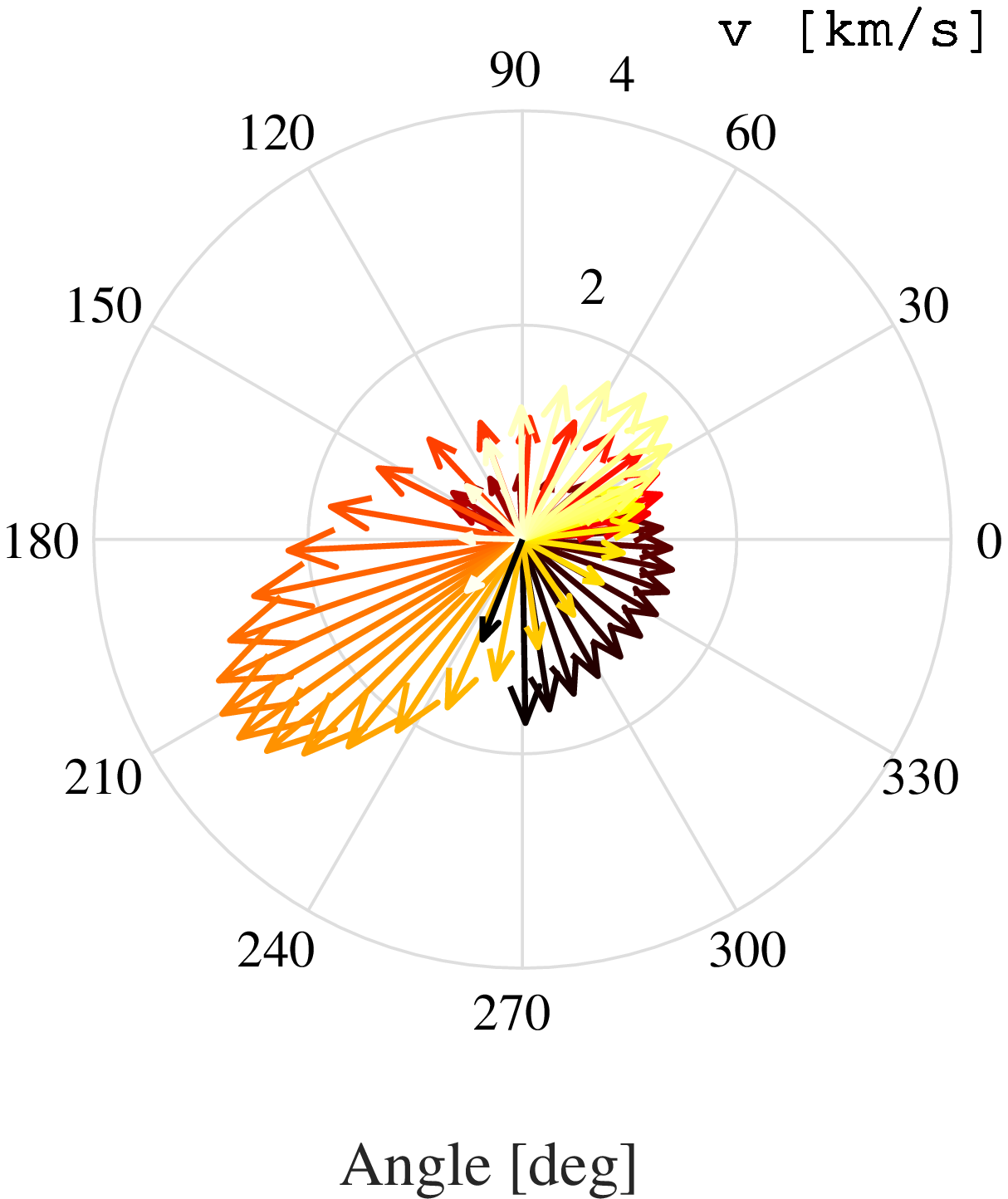}
   \caption{Compass polar diagrams of the horizontal velocity vector of six magnetic elements chosen among the 35 analysed in this work. The color scale represents the temporal evolution (from dark red to light yellow). The azimuthal lines (latitudes) indicate the magnitude of the horizontal velocity vector in $km/s$. }
    \label{Fig:compass}
   \end{figure*}
The observation started at 07:57:39 UT and lasted for $47$ minutes with a cadence of the spectral scans of $28$ s ($100$ spectral scans). The pixel scale was $0.034$ arcsec/pixel for the Ca II H filter. 
The spatial resolution of the images is equivalent to $\sim120$ km in the solar photosphere.
The standard calibration procedure including the MOMFBD \citep[Multi-Object Multi-Frame Blind Deconvolution,][]{MSnoort05} restoration aimed at limiting seeing-induced distortions in the images. \\
In Fig. \ref{Fig:maps}, we show a typical example of images in the core of the Ca II H line (upper panel), where the selected magnetic elements are highlighted by yellow circular symbols. In the same Figure (lower panel), we also show a photospheric image of the same field of view taken at $395.3$ nm ($1$ \AA~ bandpass) .\\
The identification of the chromospheric features was performed on each Ca II line core image of the available series by applying a procedure based on determining the center of mass of the intensity distribution in windows of area $10\times10$ pixel$^2$ encompassing the intensity enhancement co-spatial to the photospheric feature. For more detailed information on the identification we refer the reader to Paper I.
At each time step, the two components of the horizontal velocity of each identified magnetic element were determined as the time derivatives of the measured position of the feature.\\

\subsection{Empirical mode decomposition}
In order to isolate the low frequency evolution of the horizontal velocity vector and to filter out the high frequency noise that can affect the signals, we applied the method of empirical mode decomposition (EMD). EMD was introduced by \citet{huang1998empirical} as a technique for the regularization of a signal before the application of Hilbert transform. EMD decomposes a signal into a set of intrinsic mode functions (IMFs), which represent different oscillations at a local level. For a complete  introduction to EMD and its application to solar physics data, we refer the reader to \citet{2004ApJ...614..435T}. Another example of a solar application involving MHD waves is by \citet{2011ApJ...729L..18M}.\\
In contrast to the Fourier method, that applies to rigorously stationary and linear time series, the characteristic time-scales of the IMFs preserve all the non-stationarities of the signal as well as its non-linearities. Indeed, an IMF is defined as a local mode, which satisfies the following conditions: \textit{i)} the number of extrema and zero-crossing should be equal or differ at the most by one, \textit{ii)} at any time the mean value of the upper and lower envelopes, defined from the local maxima and minima, respectively, is zero. It is important to point out that the EMD technique does not require any \textit{a priori} assumption, the decomposition being based on the data itself.
As a result of applying the EMD technique, the original signal $v(t)$ is decomposed into a set of IMFs and a residue $R$, so that one can write:
\begin{equation}
v(t)= \sum_{i=1}^n IMF_{i}(t)+R(t).
\end{equation}
It is important to underline that the EMD method, based on the extraction of the energy contained in the intrinsic time scales of the signal, can be applied successfully to non-stationary signals. These properties make the method very attractive for solar MHD wave research. \\
It has to be noted that, in contrast to Fourier transforms, EMD does not require any transformation of the signal, thus preserving the original non-linearities (if any) of the process.\\
The EMD technique has been already employed to study kink waves in small-scale magnetic elements and more information about the application of this technique can be found in \citet{2014A&A...569A.102S}.\\
Distinct from earlier works available in the literature, in this study we examine the slow temporal changes of the orientation of the velocity vector of the magnetic elements, instead of the oscillations of its magnitude. To do this, we apply EMD to both components of the horizontal velocity of each magnetic element investigated, and extract the low frequency part of the signal as in Fig. \ref{Fig:EMD}. Each velocity signal is decomposed into $5$ IMFs. The first IMF containing the high frequency part of the signal is then neglected, while the following ones are used to reconstruct the signal itself. Indeed, the first IMF mostly captures noise \citep{flandrin2004detrending}, and high-frequency perturbations due for instance to intergranular turbulence \citep{2014A&A...563A.101J}.

\section{Results}
\subsection{A case study}
Before showing the results derived from the analysis of the $35$ magnetic elements selected, in this section we analyse in detail a case study in order to highlight the key points. This is also done to better describe the methods used, and provide a more detailed insight into the results for a particular yet representative case. In this regard, we selected the magnetic element represented by a red star in Fig. \ref{Fig:maps}. This SSME is tracked with an automated procedure that tracks the center-of-mass of the intensity distribution of the element itself. After the automated procedure the tracking is verified by visual inspection, as for all other SSMEs studied in this work. The trajectory $\textbf{s}(t)$ of the selected SSME in time is shown in Fig. \ref{Fig:elem24}. Superimposed on the trajectory of the SSME, we plot the velocity vector obtained from the derivative of $\textbf{s}(t)$. The horizontal velocity is characterized by a broad spectrum of oscillations (see the right panel in Fig. \ref{Fig:elem24}), from $\sim 1-2$ mHz up to $\sim 10$ mHz. This is the case for both components of the horizontal velocity vector, which display a good agreement. In the same panel we also overplot the coherence (smoothed with an averaging window 3 points wide) between the two spectra defined by:
\begin{equation}
C(\nu)=\frac{|P_{xy}(\nu)|^{2}} {P_{xx}(\nu)P_{yy}(\nu)},
\end{equation}
where $P_{xy}(\nu)$ is the cross spectral density between $v_{x}$ and $v_{y}$ (the two components of the horizontal velocity vector $\dot{\vec{s}}(t)$), and $P_{xx}(\nu)$ and $P_{yy}(\nu)$ the power spectral densities of $v_{x}$ and $v_{y}$, respectively.\\
It is interesting to note that the visual inspection of the trajectory itself already gives the impression of a rotation of the displacement vector. This appears more evident in Fig. \ref{Fig:elem24compass}, where we plot the EMD filtered horizontal velocity vector $\dot{\vec{s}}(t)$ in polar coordinates. The color scale in this (and any further) polar plot encodes the temporal evolution (from dark red to light yellow). These plots shows that the orientation of the velocity vector $\dot{\vec{s}}(t)$ changes smoothly in time and does not present jumps in the orientation itself. In other words, there exists a long-term memory of the process, which determines the evolution of the velocity vector. This can be also seen in the right panel of the same figure, where we plot the same information contained in the polar plot previously described, but unrolled it along the time axis. This graph clearly displays a rotation of the velocity vector in time. The rotation of $\dot{\vec{s}}(t)$ takes place over the first $30$ min only, while in the remaining fraction of the lifetime of the SSME, no rotation of the velocity vector is observed. Referring to the trajectory of the SSME shown in the upper right panel of Fig. \ref{Fig:elem24}, we see that the first part of the element lifetime is marked by a distinguishable helical motion of the flux tube (trajectory points lying in the right half part of the plot). In this regard, the polar maps of Fig. \ref{Fig:elem24compass} offer a much clearer visualization of this behaviour, thus in the rest of this work we will focus on such polar plots of the velocity vector. This process is schematically depicted in Fig. \ref{Fig:cartoon}.\\
The results in Figs. \ref{Fig:elem24}, \ref{Fig:elem24compass}, and \ref{Fig:compass} show that the rotation of the horizontal velocity vector is a temporally coherent process that occupies a significant fraction of the lifetime of the selected magnetic element ($\sim 30$ min). Indeed, the orientation of the velocity vector is not randomly distributed in space, but follows a helical evolution that suggests a phase lag between $v_{x}$ and $v_{y}$.

\subsection{Analysis of the SSMEs sample}
In Fig. \ref{Fig:compass}, we visualize the evolution of the horizontal velocity vector in polar coordinates (after EMD decomposition) of a sample of 6 magnetic elements selected among the 35 elements. The compass plots reveal that the horizontal velocity oscillations of the magnetic elements are not randomly oriented in space, but follow nearly helical trajectories (i.e. the velocity vector evolves smoothly in time, without sudden changes of its orientation). In other words, the horizontal velocity of the magnetic elements results to be elliptically polarised, for a significant fraction of the SSMEs lifetime.  
In addition to this, the helical motion of the velocity vector is sometimes seen to revert the direction of its angular motion from clockwise to counterclockwise and viceversa).\\
\begin{figure}
  \centering
  \includegraphics[trim=0cm 0cm 0cm 0cm, clip, width=7.cm]{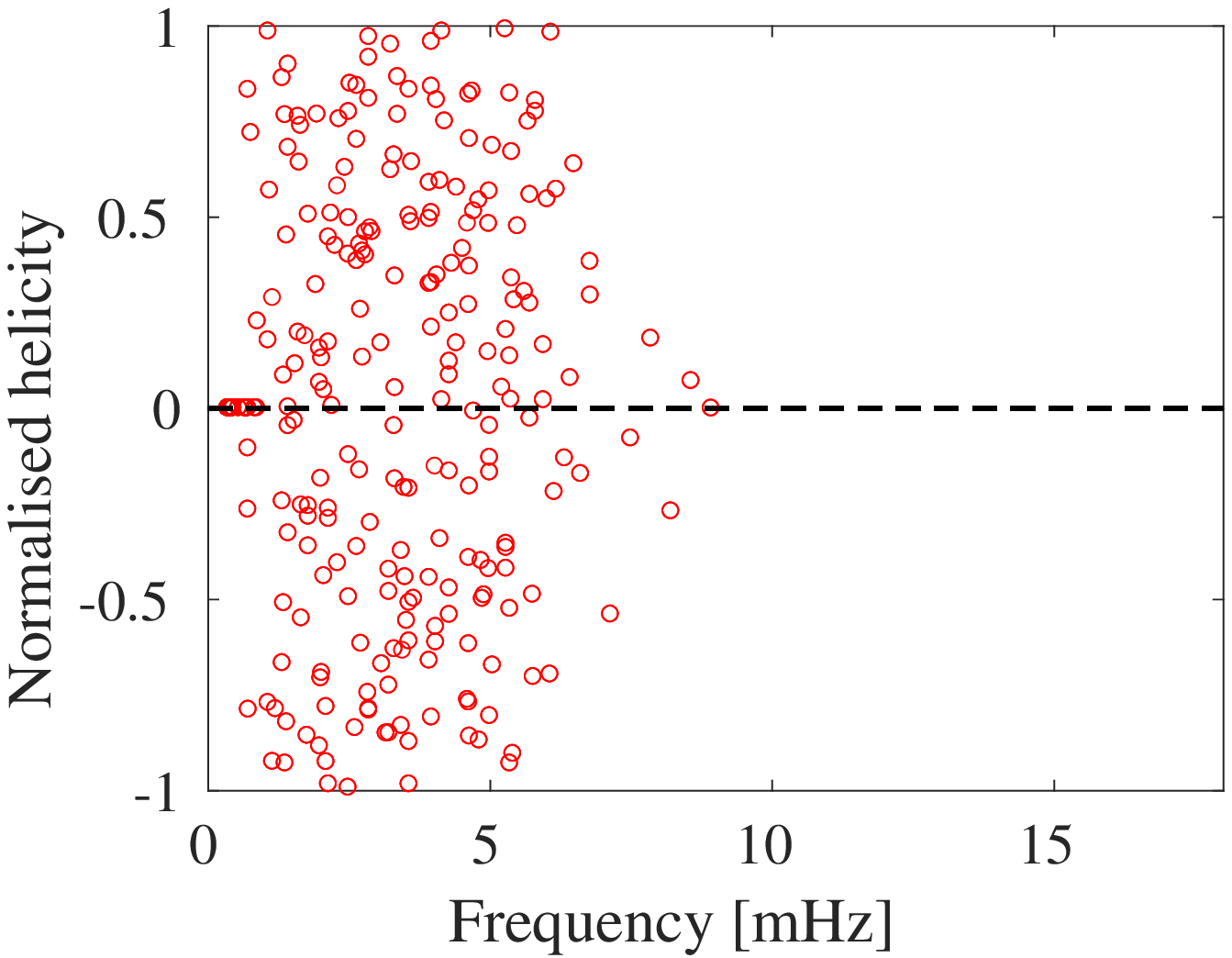}
  \includegraphics[trim=0cm 0cm 0cm 0cm, clip, width=7.cm]{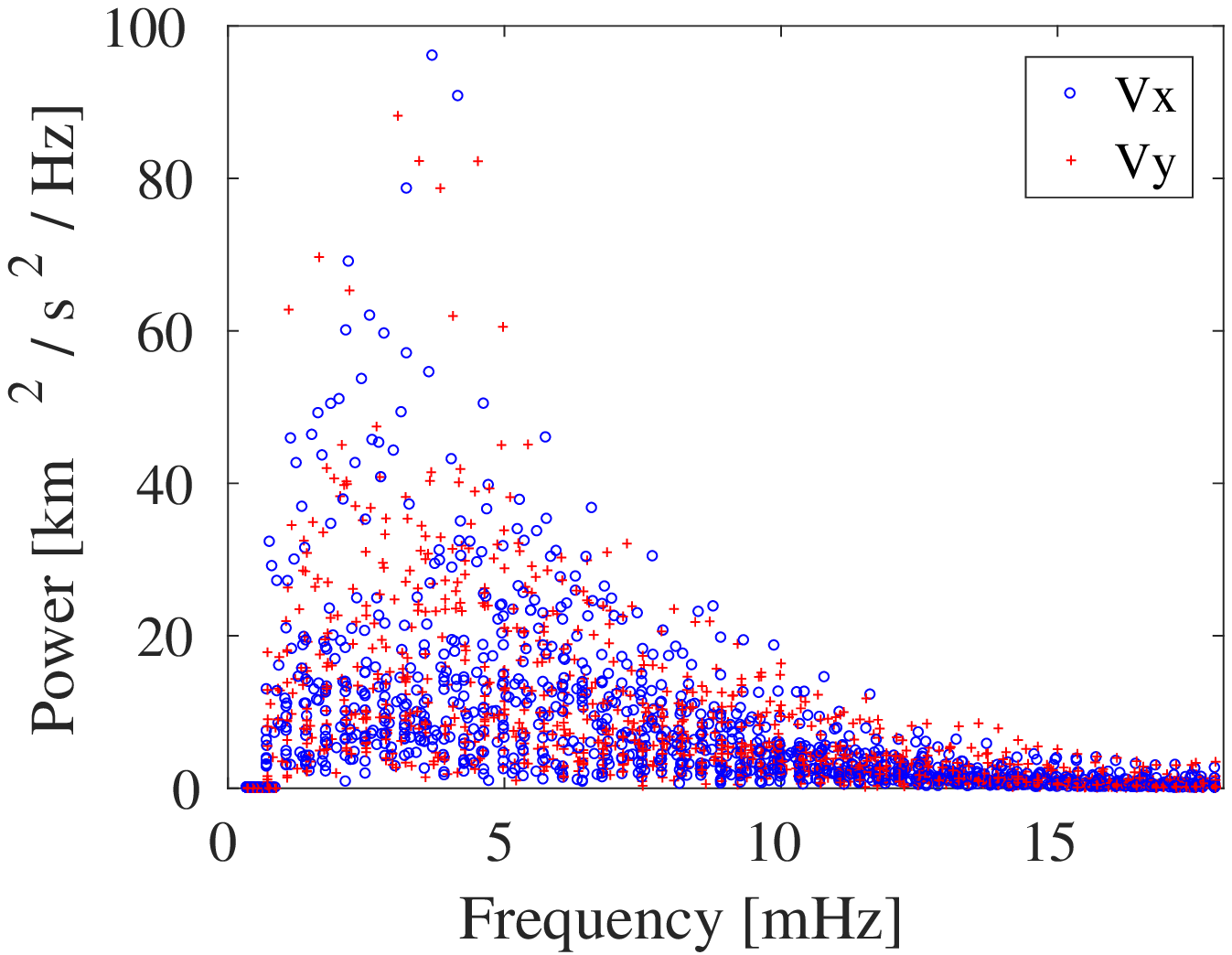}
  \includegraphics[trim=0cm 0cm 0cm 0cm, clip, width=7.cm]{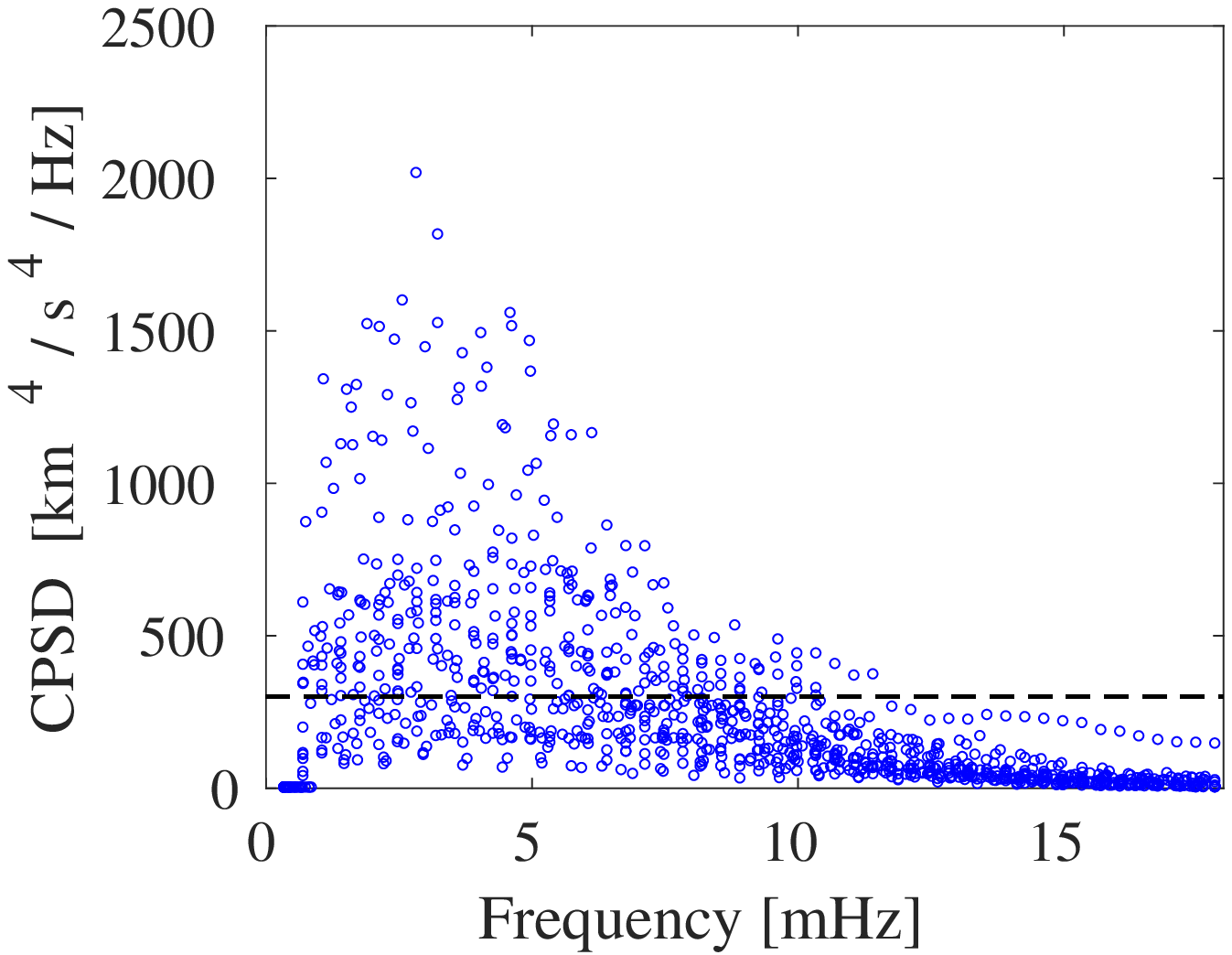}
   \caption{\textit{Top:} Helicity spectrum of the horizontal velocity vector of the 35 magnetic elements investigated. Only those Fourier modes with simultaneously a coherence larger than $0.8$, and a cross-correlation larger than the threshold indicated in the bottom panel of this figure. \textit{Middle:} Power spectra of the two components of the horizontal velocity for all the elements selected. Only modes with coherence larger than 0.8 are shown. \textit{Bottom:} Cross-power spectral density computed between the two components of the velocity vector. Only modes with coherence larger than 0.8 are shown. The horizontal dashed line represents the threshold used in the top panel of this figure.}
    \label{Fig:helicity}
   \end{figure} 
In order to give an independent and more quantitative characterization of the elliptical motion of the velocity vector seen in the examples of Fig. \ref{Fig:compass}, we estimate the helicity $H(\omega)$ of the velocity perturbations. The helicity can be written as follows \citep{1997ApJ...488..482C}:
\begin{equation}
H(\omega)=\frac{2~Im[a_{x}^{*} ~ a_{y}^{*}]}{\omega},
\end{equation}
where $\omega$ is the frequency, and $a_{x}$ and $a_{y}$ the FFT transforms of $v_{x}$ and $v_{y}$, respectively.\\
Let us now normalise the helicity as follows:
\begin{equation}
\sigma(\omega)=\frac{\omega H(\omega)}{ |a_{x}|^{2} + |a_{y}|^{2}}.
\end{equation}
It is worth noting that although the helicity was initially used to study fluctuations in the solar wind, it is clear that eq. 3 can be applied to any time series, as it represents a relation between modes of the Fourier space.
The normalised helicity of velocity oscillations  of all the 35 magnetic elements is shown in Fig. \ref{Fig:helicity} (upper panel). The figure shows that a large number of modes in the spectrum display a $\sigma>0$ or $\sigma<0$. This means, clockwise or counter-clockwise rotation, respectively, and confirms the earlier derived polarization of the horizontal velocity vector of the magnetic elements in the solar chromosphere. In the same plot, we only show those points with coherence larger than 0.8, and cross-power spectral density (CPSD) larger than $300~km^{4}s^{-4}Hz^{-1}$ (see bottom panel of the same figure). This threshold is chosen in order to isolate the most prominent peak in the cross-correlation plot in the range $0-10$ mHz. The large coherence also represents a high confidence level ensuring the reliability of the measurements, with a value of 0.8 being a very stringent confidence threshold.

\subsection{A toy model}
\begin{figure}
  \centering
  \includegraphics[trim=0cm 0cm 0cm 0cm, clip, width=7.cm]{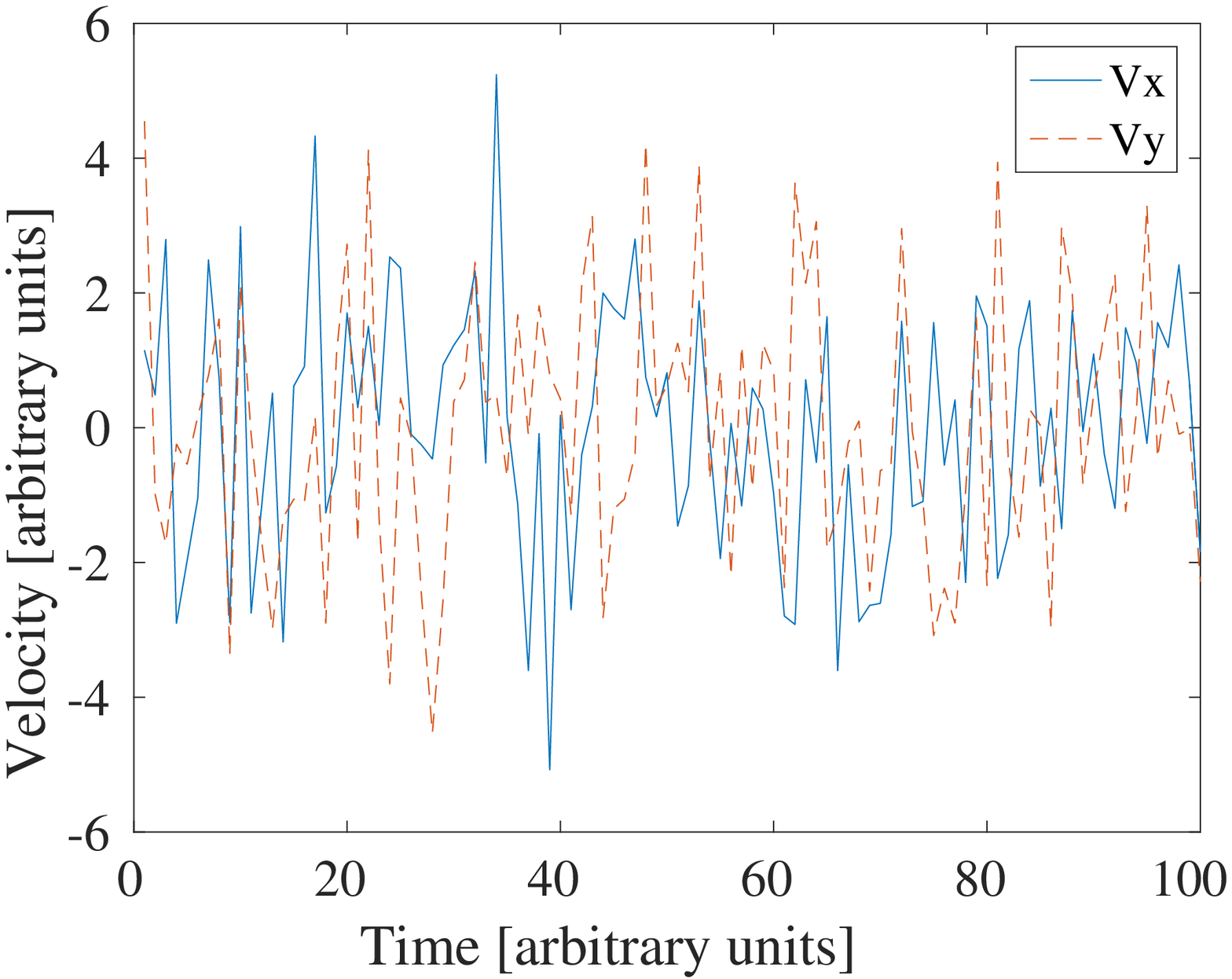}
  \includegraphics[trim=0cm 0cm 0cm 0cm, clip, width=7.cm]{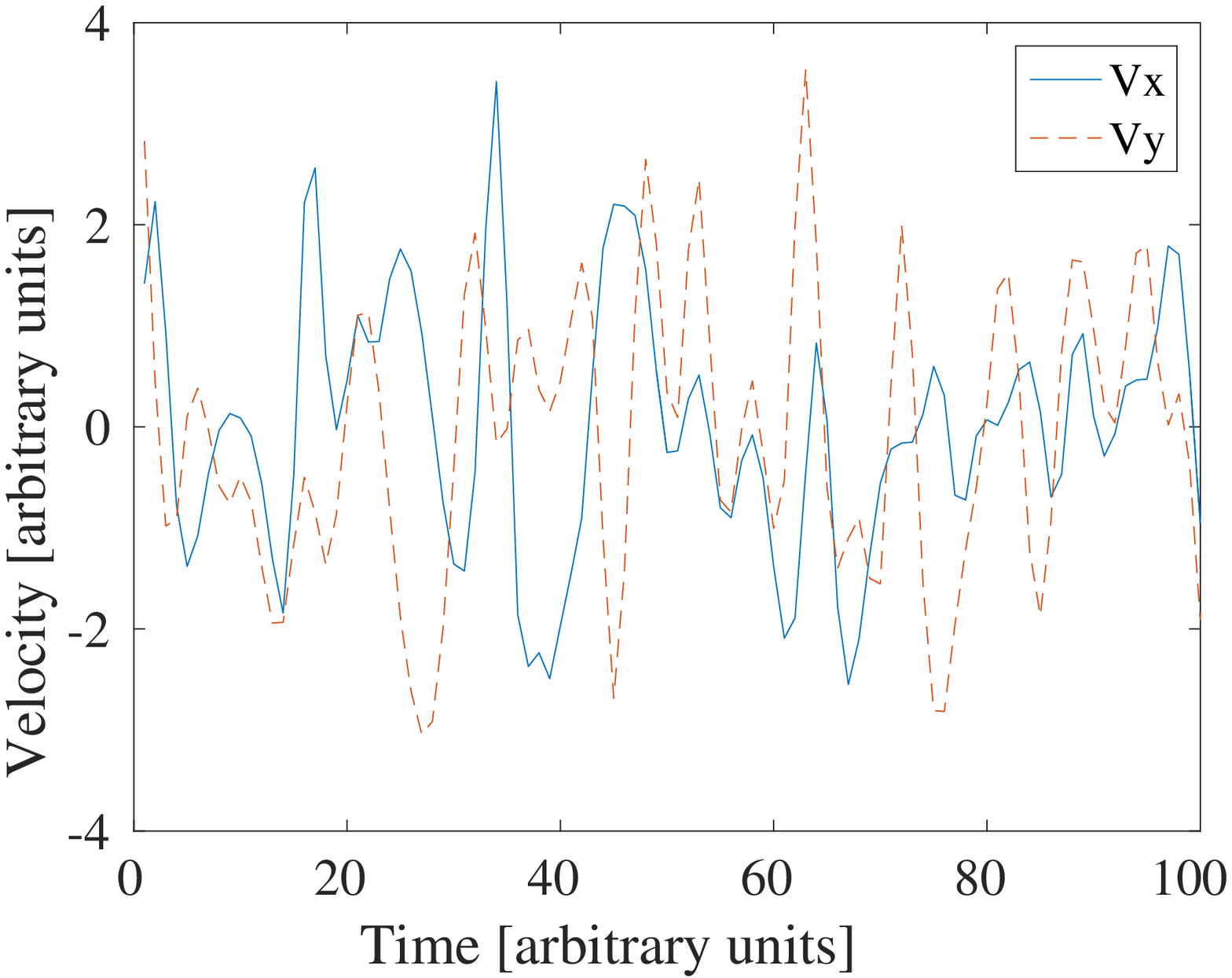}
  \includegraphics[trim=0cm 0cm 0cm 0cm, clip, width=7.cm]{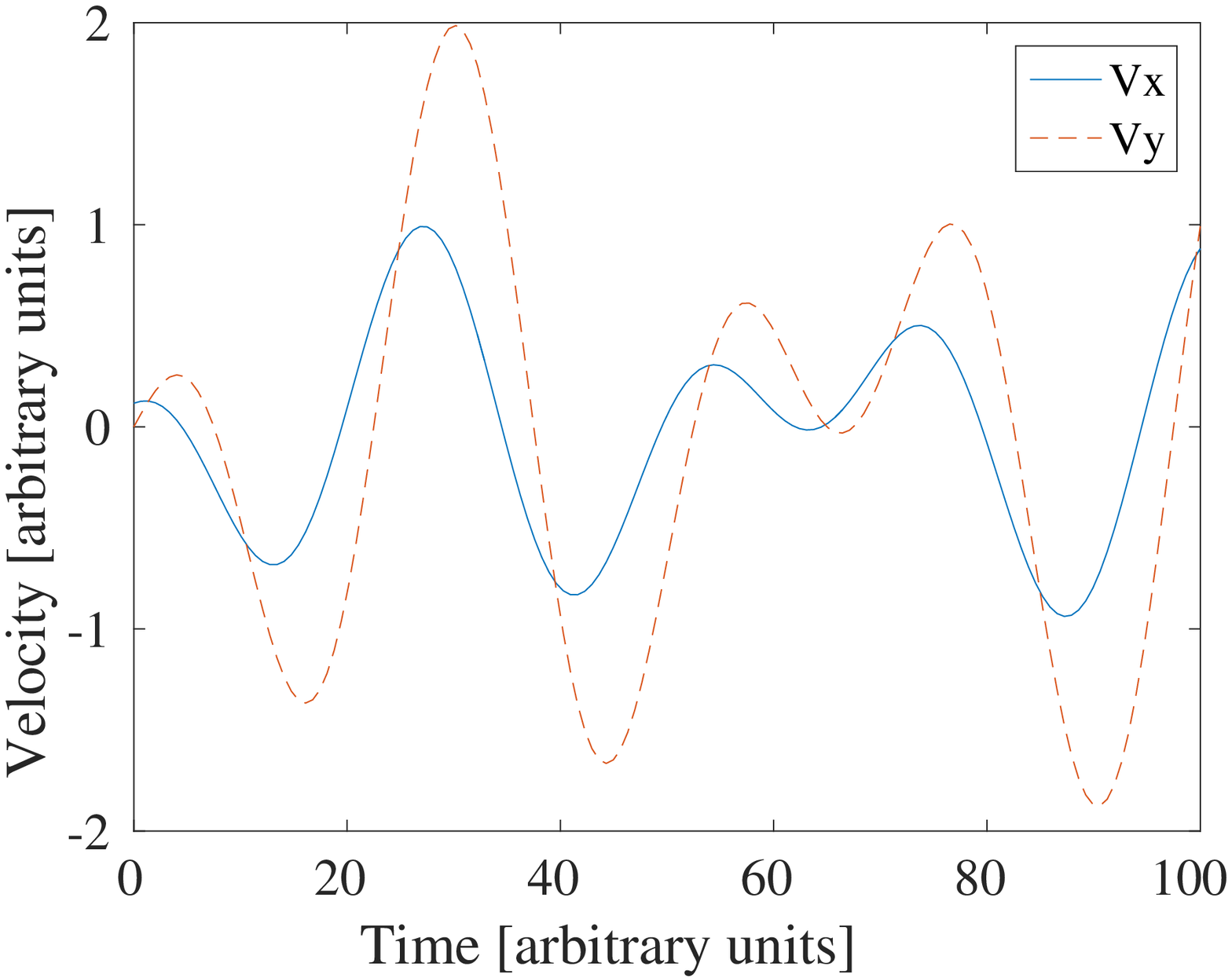}
   \caption{\textit{Top panel }: simulated random velocity signal with normal distribution. \textit{Middle panel:} same velocity signal filtered using the EMD. \textit{Bottom panel:} simulated velocity signal made with two superimposed pulses in the x and y direction respectively. The two pulses are exactly the same, but slightly shifted and with different amplitudes.}
    \label{Fig:random_proc}
   \end{figure}

\begin{figure}
  \centering
  \includegraphics[trim=0cm 0cm 0cm 0cm, clip, width=6.cm]{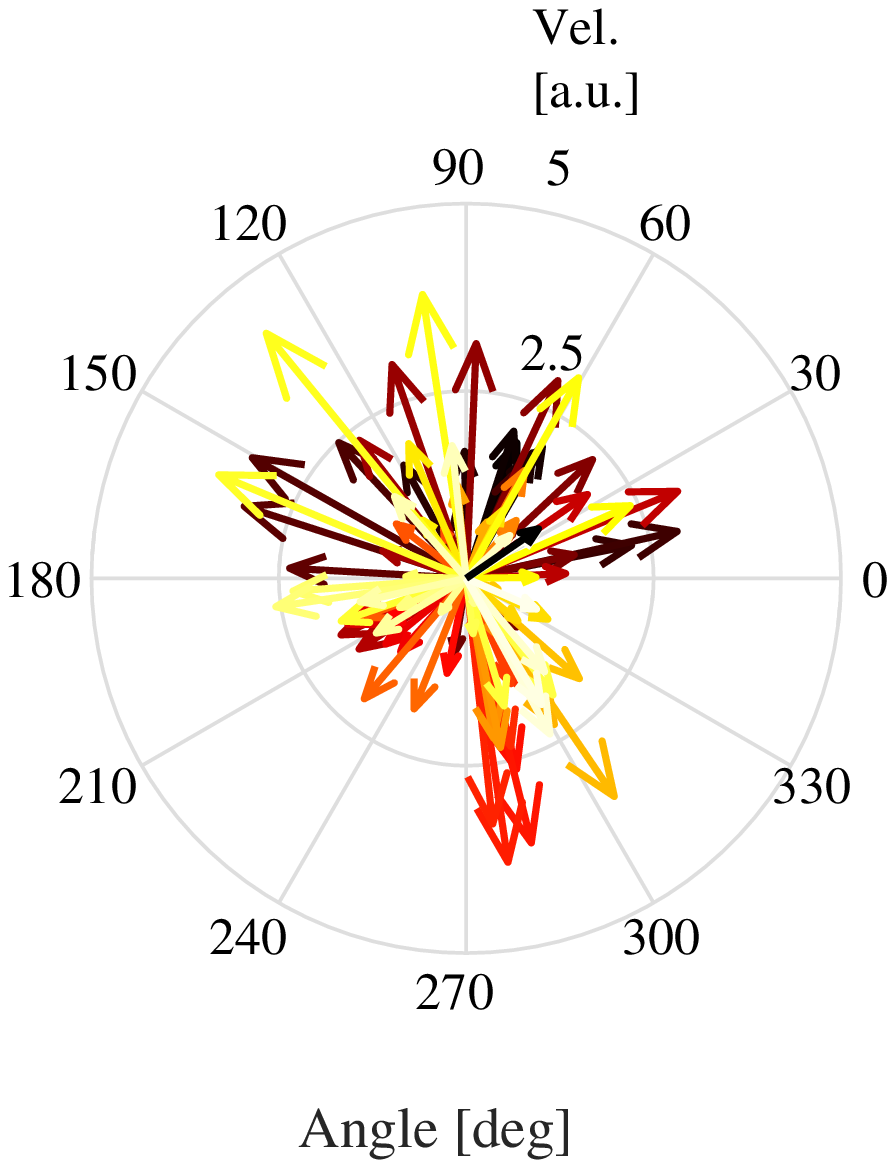}\\
  \vspace{1cm}
  \includegraphics[trim=0cm 0cm 0cm 0cm, clip, width=6.cm]{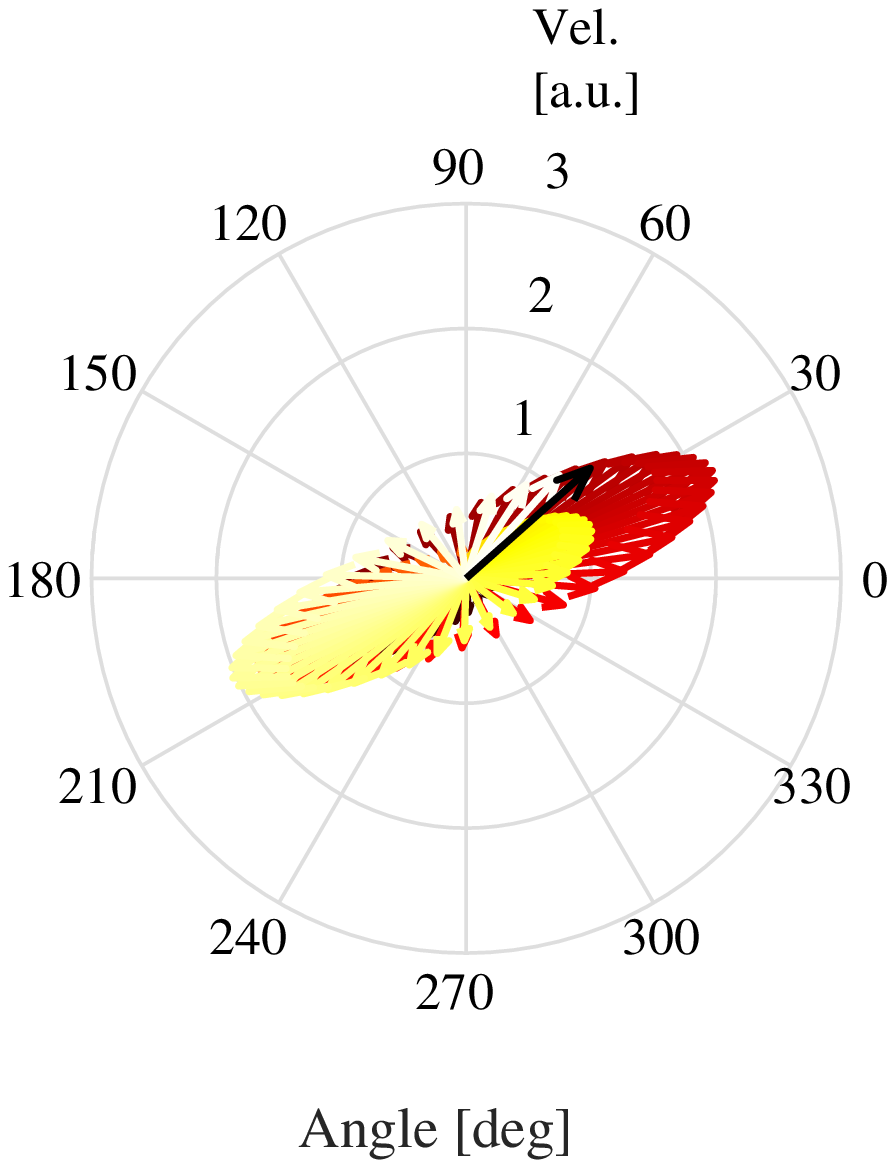}
  \caption{Compass diagram of the simulated random velocity signal with normal distribution (top), and the one obtained from a superposition of two velocity pulses in two perpendicular directions (bottom).}
    \label{Fig:compass_sim}
   \end{figure}

With the aim of validating our results, we applied our method to two control simulations: a random process, and an oscillatory process, in which two pulses with different directions are superimposed. \\
In the first case, a random velocity with normal distribution is simulated, see upper panel of Fig. \ref{Fig:random_proc}. 
In order to apply exactly the same method used with the real data, we filtered the simulated velocity signal employing the EMD (middle panel of Fig. \ref{Fig:random_proc}). The polarization of this filtered signal is then visualized through a compass diagram just like in the case of the observed data. Besides, we simulated a simple oscillatory signal of the form:
\begin{equation}
v_{x,y}(t)=A sin \frac{2 \pi t}{T_{1}} sin \frac{2 \pi t}{T_{2}},  
\end{equation}
where $A$ is the amplitude of the velocity signal, $T_{1}$ is period, and $T_{2}$ is long-term modulation to simulate a wave train (see lower panel of Fig. \ref{Fig:random_proc}).\\
The compass diagrams of the two simulated velocity signals are shown in Fig. \ref{Fig:compass_sim}. In comparison with the oscillatory signal, the random process does not display polarization (top panel). This confirms the validity of our findings. In contrast, the same kind of polarization of the compass diagram is indeed obtained with a superposition of two pulses in different directions (bottom panel).

\section{Discussions and conclusions}
The results presented above show low frequency ($<5-6$ mHz) kink-like horizontal oscillations of SSMEs in chromosphere with an elliptic polarization of the velocity vector (see Fig. \ref{Fig:cartoon} for a schematic representation) are observed in all the 35 SSMEs analysed in the quiet chromosphere.  
This frequency range is inconsistent with residual seeing aberrations. In this regard, it is worth noting that the orientation of the velocity vector in time is not random, but reflects a long-term coherency (several minutes). This timescale is by far longer than the typical timescale of atmospheric turbulence (a few ms). Indeed \citet{2016SPIE.9909E..7PS} have deeply analysed the de-correlation time scales of AO residual aberrations using on-sky data, and found them to be of the order of $10-20$ ms at visible wavelengths ($\sim 630$ nm). This is consistent with independent works in literature \citep[see for example][]{1996PASP..108..456D}. Since the turbulence timescales at the Ca II H wavelengths is even shorter (\citet{roddier1999adaptive}), we expect a large reduction of the turbulence coherence time, resulting in a turbulence timescale that is even more inconsistent with that of the process observed here. In addition, since the seeing residual aberrations are a random process, one can not expect any coherent helical motion of the velocity vector from them.\\
We would like to comment here on the selection of the sample of SSMEs used in this work. Indeed, the choice of the 35 chromospheric magnetic elements analysed in Paper I, and in this work, is twofold. Firstly, in the chromosphere the magnetic elements are not forced by the surrounding granular flows, thus the horizontal oscillations measured represent their free oscillation, and cannot be misinterpreted as the result of the photospheric forcing itself. Secondly, in Paper I, it was shown that these elements, observed as brightenings in Ca II H, could be unambigously associated with magnetic elements. Indeed, here it was possible to clearly locate their corresponding circular polarization signals in the photosphere. This fact, in particular, ensure that the brightenings tracked in the Ca II H data represent the chromospheric counterpart of photospheric magnetic features.  \\
A helical motion was already observed in solar spicules by e.g. \citet{2007Sci...318.1574D, 2008ASPC..397...27S}. For a complete historical review of the observations of waves in solar spicules we refer the reader to \citet{zaqarashvili2009oscillations}. In the attempt to put this observational aspect in a theoretical framework, \citet{2008ApJ...683L..91Z} have shown that the superposition of random photospheric pulses with different orientations may easily explain the observed polarised motion. Indeed, these authors demonstrated that each photospheric forcing pulse can excite a kink wave in the flux tube whose polarization plane depends on the pulse itself. The superposition of two or more pulses, and then of different kink waves polarized in different planes, may give rise to a complex oscillation of the flux tube and set up helical waves. Our observational results are perfectly in agreement with this theoretical prediction, and can be seen as the counterpart of the polarized motion in spicules observed by  \citet{2007Sci...318.1574D}, and predicted by \citet{2008ApJ...683L..91Z}. Moreover, our results confirm that the superposition of different driving pulses, with different amplitudes, are effective in generating kink waves with elliptical polarisation that can propagate from the photosphere to the chromosphere, where they still maintain their kink-like oscillatory behaviour.\\
In this regard, \citet{zaqarashvili2009oscillations} noted that the flux tube expansion in the solar chomosphere may hamper the propagation of kink waves. However, our results show that, as soon as the heights spanned by the Ca II H are concerned ($\sim 700$ km above the photosphere), the kink wave appears the most plausible explanation to the observed oscillations \citep{2007Sci...318.1572E}. \\

\acknowledgments
This work is partly supported by the "Progetti di ricerca INAF di Rilevante Interesse Nazionale" (PRIN-INAF) 2014 and PRIN-MIUR 2012 (prot. 2012P2HRCR) entitled "Il sole attivo ed i suoi effetti sul clima dello spazio e della terra" grants, funded by the Italian National Institute for Astrophysics (INAF) and Ministry of Education, Universities and Research (MIUR), respectively. This study received funding from the European Unions Seventh Programme for Research, Technological Development and Demonstration, under the Grant Agreements of the eHEROES (n 284461, www.eheroes.eu), SOLARNET (n 312495, www.solarnet-east.eu). We also thank Luc Rouppe van der Voort, and Alice Cristaldi for their help in supporting the data reduction. RE is grateful to STFC (UK), the Royal Society and the Chinese Academy of Sciences Presidents International Fellowship Initiative, Grant No. 2016VMA045 for support received. SJ receives support from the Research Council of Norway.

\bibliographystyle{yahapj}
\bibliography{lib.bib}

\end{document}